\documentclass[aps,twocolumn]{revtex4}
\usepackage{amsmath,amssymb}
\usepackage{graphics,graphicx}
\usepackage{dcolumn,bm}

\begin{document}
\title{Kolkata Restaurant Problem as a generalised El Farol Bar Problem}

\author{Bikas K. Chakrabarti}
\email{bikask.chakrabarti@saha.ac.in}
\affiliation{Theoretical Condensed Matter Physics Division and
Centre for Applied Mathematics and Computational Science,
Saha Institute of Nuclear Physics
1/AF Bidhan Nagar, Kolkata 700064, India.}


\begin{abstract}  
Generalisation of the El Farol bar problem to that of many bars
here leads to the Kolkata restaurant problem, where the decision to go
to any restaurant or not is much simpler (depending on the previous
experience of course, as in the El Farol bar problem). This
generalised problem can be exactly analysed in some limiting
cases discussed here.
The fluctuation in the restaurant service can be shown to have
precisely an inverse cubic behavior, as widely seen in the
stock market fluctuations.
\end{abstract}


\maketitle
\section{Introduction}

The observed corrlated fluctuations in the stock 
markets, giving power law tails for large fluctuations (in contrast to the
traditionally assumed  
exponentially decaying Gaussian fluctuations
of random walks), were schematically incorporated
 in the El Farol Bar problem of Arthur~\cite{BKC:Arthur:1994}. 
Here  the decision to occupy the bar (buy the stock) or to remaing at home 
(sell the stock) depends on the previous experience of the ``crowd" exceeding 
the threshold (of pleasure or demand level) of the bar (stock), and the 
strategies available. The resulting Minority Game 
models~\cite{BKC:ChalletMarsiliZhang:2005} still fail 
to get the ubiquitus inverse cubic law of stock 
fluctuations~\cite{BKC:MantegnaStanleyEconophys:1999}. In the Fiber 
Bundle models~\cite{BKC:LNP705:2006} of materials' fracture, 
or in the equivalent Traffic Jam  models~\cite{BKC:Chakrabarti:2006}, 
the fibers or the roads fail due to load, exceeding the
(preassigned)  random 
threshold, and the extra load gets redistributed in the surviving fibers or 
roads; thereby inducing the corelations in the fluctuations or ``avalanche" 
sizes. The avalanche distribution has a clear inverse cubic power law tail
in the ``equal load sharing" or ``democratic" fiber bundle 
model~\cite{BKC:Pradhan:2002,BKC:Pradhan:2007}. 

In the El Farol Bar problem~\cite{BKC:Arthur:1994}, 
the Santa Fe people decide whether to go 
to the bar this evening, based on his/her experince last evening(s). 
The bar can roughly
 accommodate half the poulation of the (100-member strong) Institute and
the people coming
to the bar still enjoy the music and the food. 
If the crowd level goes beyond this level,
people do not enjoy and each of those who came to the bar thinks that they
would have done better if they
stayed back at  home! Clearly, people do not randomly choose to
come to the bar or stay at home (as assumed in a random walk model); 
they exploit their past experience and their
respective strategy (to decide on the basis of the past experience).  
Of course the people here are assumed to
 have all the same informations at any time (and 
their respective personal experiences) available
to decide for themselves independently and parallely; they 
do not organise among themselves and go to the bar! 
Had the choice been completely random, the occupation fluctuation of the bar
would be Gaussian. Because of the processes involved in deciding to go or not,
depending on the respective experience, the occupation 
fluctuation statistics  changes. The ``minority" people win
such games (all the people ``lose" their money if the bar gets ``crowded",
or more than the threshold, say, 50 here); the bar represents either the
``buy" or ``sell" room and  the (single) stock  
fluctuations are expected to be
represented well by the model. The memory size and the bag of tricks for each
agent in the Minority Game model made this  process and the resulting
analysis very precise~\cite{BKC:ChalletMarsiliZhang:2005}. 
Still, as we mentioned earlier, it cannot explain the origin of the
ubiquitous inverse cubic law of fluctuations (see 
e.g.~\cite{BKC:MantegnaStanleyEconophys:1999}).

We extend here this simple bar problem to many bars (or from single 
stock to many), and define 
the Kolkata Restaurant problem. The 
number of restaurants in Kolkata, unlike
in Santa Fe, are huge. Their (pleasure level) thresholds are also widely
distributed. The number of people, who choose among these restaurants, are 
also huge! Additionally, we assume that the decision making part here in 
the Kolkata Restaurant problem to be extremely simple 
(compared to the El Farol bar problem): 
if there had been any ``bad experience" 
(say, crowd going beyond threshold level) 
in any restaurant any evening, all those who came
there that evening avoid that one for a
suitably chosen long time period ($T$) and starting next evening 
redistribute this extra
 crowd or load equally 
among  the rest of the restaurants in Kolkata
(equal or democratic crowd sharing).
This restaurant or the stock 
fails (for the next $T$ evenings). As mentioned before, this failure 
will naturally increase the crowd level in all the other restaurants,
thereby inducing the possibility of further failure of the restaurants 
in service (or service
stocks). If $T$ is finite but large, the
system of restaurants in Kolkata would actually organise itself into a 
``critical" state with  a robust and precise inverse cubic power law 
of (occupation or in-service number) fluctuation. This we will show here
analytically.

\section{Model}

\begin{figure}
\centering{
\includegraphics[width=7.0cm]{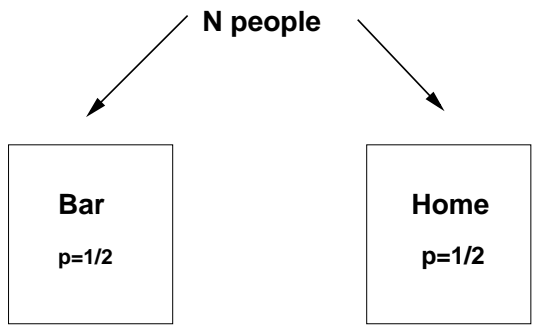}
}
\caption{\label{CCFig1}
El Farol bar problem: To go (buy) or not to go (sell) to a single bar 
(stock). Each of the $N$ people have a choice to stay in their respective
homes (collectively represented by a single `Home' here) or go to
the bar each evening. The bar has a pleasure threshold
$N/2$ ($p=1/2$) beyond which people get disappointed (lose the game)
and the `minority' deciding to stay at home that evening win. 
In the reverse case, the bar people become minority and each of them win.
}\end{figure}
\begin{figure}
\centering{
\includegraphics[width=7.0cm]{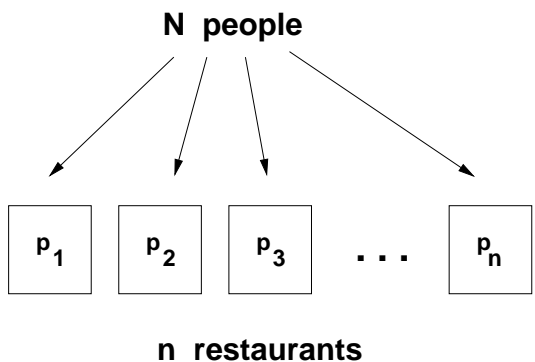}
}
\caption{\label{CCFig2}
Kolkata Restaurant problem: Many choice (stock) problem.
$N$ people in the city, each has the same choice
to go to any of the restaurants (stocks) and therefore, each retaurant gets
$p=N/n$ fraction of the crowd to start with. 
If $p > p_i$ on any evening, then the $p$ fraction of the people going
to that $i$th restaurant gets dissatisfied, and the $i$th restaurant
(stock) falls out of choice for the next $T$ evenings. 
$p$ then increases to $N/(n-1)$ and that may lead to a further
failure of the $j$th restaurant (if $p_j < N/(n-1)$) and so on.
}\end{figure}

Let $n$ represent the number of restaurants in Kolkata. They are certainly not
identical in their size and comfort threshold levels. let $p_1, p_2, ...,
p_n$ denote respectively the crowd threshold levels of these $n$ restaurants.
If, in any evening, the crowd level  $p$ in the $i$th 
restaurant exceeds $p_i$, then all the $p$ number (fraction, when normalised)
of persons coming to the $i$th restaurant that evening decide
not to come to that restaurant for the next $T$ evenings, and the 
$i$th restaurant goes out of service for the next $T$ evenings
(because others got satisfaction from the restaurants they went last
evening, and therefore do not change their choice). If $N$ is the
total number (assumed to be large and fixed) of people regularly 
going to various restaurants in Kolkata, and if we assume that people
choose completely randomly among all the  restaurants in service (democratic 
or equal load sharing hypothesis and 
``knowledge" of the ``in-service" restaurants
available to everybody), the ``avalanches" dynamics of these restaurants
to fall out of service,
can be analytically investigated if $T \to \infty$ and the threshold
crowd level distribution $\rho(p_i)$  for the restaurants are 
known (see Figs.~\ref{CCFig1} and~\ref{CCFig2}).

\section{Avalanche Dynamics: Infinite $T$}

This avalanche dynamics can be represented by recursion
relations in discrete time steps. 
Let us  define $U_t(p)$ to be the fraction of in-service restaurants in 
the city
that survive after (discrete) time step $t$ (evenings), 
counted from the time $t=0$
when the load or crowd  (at the level $p = N/n$) is put in the system
 (time step indicates the number of crowd 
redistributions). As such, $U_t(p =0)=1$ for all $t$ and $U_t(p)=1$
for $t=0$ for any $p$; 
$U_t(p)=U^*(p) \ne 0$ for $t \to \infty$ if 
$p < p_c$, and 
$U_t(p)=0$ for $t \to \infty$ if $p > p_c$.
\begin{figure}
\centering{
\includegraphics[width=7.0cm]{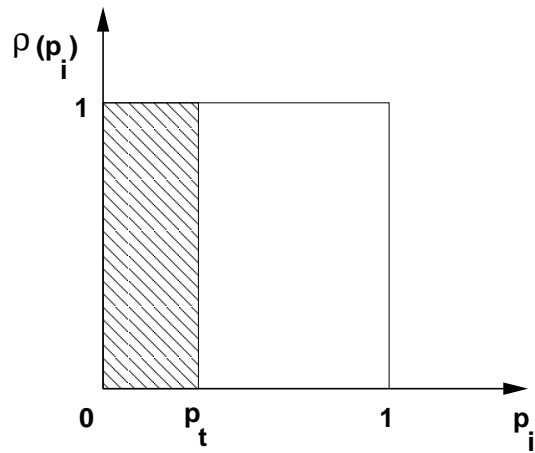}
}
\caption{\label{CCFig3}
Density $\rho(p_i)$ of the crowd handling capacities $p_i$ of the
restaurants. It is assumed here to be uniform upto a threshold value
(normalised to unity). At any evening $t$, if the crowd level is $p_t$,
restaurants having $p_i \le p_t$ all fail and the fraction
$1-p_t$ of restaurants remain in service after that evening.
}\end{figure}

If $p$ is measured in the unit of the crowd threshold of the biggest
restaurant in Kolkata, or in other words, if $p$ is normalised to unity and
$\rho (p_i)$ is assumed to be uniformly distributed as shown in 
Fig.~\ref{CCFig3} (and $T \to \infty $ as mentioned), then
$U_{t}(p)$ follows a simple recursion relation 
(cf.~\cite{BKC:LNP705:2006,BKC:Chakrabarti:2006}) 
\begin{equation*}
U_{t+1}= 1-p_t;\ \ p_t = \frac{N}{U_t n}
\end{equation*}
\begin{equation}
\label{cc:recur}
{\rm or,} \ \ U_{t+1}=1-\frac{p}{U_{t}}.
\end{equation}
In equilibrium \( U_{t+1}=U_{t}=U^{*} \) and thus (1) is quadratic in
$U^{*} $ : 
\begin{equation*}
U^{*^{2}}-U^{*}+p =0.
\end{equation*}
The solution is
\begin{equation*}
U^{*}(i )=\frac{1}{2}\pm (p_{c}-p )^{1/2}; \ p_{c}=\frac{1}{4}.
\end{equation*}
Here \( p_{c} = N_c/n \) is the critical value of 
crowd level (per restaurant)
beyond which the system of (all the Kolkata) restaurants fails completely. 
 The quantity
$U^{*}(p)$ must be real valued as it has a physical meaning:
it is the fraction of the restaurants that remains in service under
a fixed crowd or load  $p$ when the load per restaurant lies
in the range \( 0\leq p \leq p_{c} \). Clearly, \( U^{*}(0)=1 \).
 The solution
with (\( + \)) sign is therefore the only physical one.
Hence, the physical  solution can be written as 
\begin{equation}
U^{*}(p)=U^{*}(p_{c})+(p_{c}- p )^{1/2};
\ U^*(p_c) = \frac{1}{2} \ {\rm and}\ p_{c}=\frac{1}{4}.
\end{equation}
For \( p > p_{c} \) we can not get a real-valued fixed
point as the dynamics never stops until \( U_{t}=0 \) when the network
of restaurants get completely out of business!

\subsection {Critical Behavior}

\subsubsection{At \(p < p_{c} \)}
It may be noted that the quantity \( U^{*}(p )-U^{*}(p_{c}) \)
behaves like an order parameter that determines a transition from
a state of partial failure of the system (at \( p \leq p_{c} \)) to a state
of total failure (at \( p > p_{c} \)) :
\begin{equation}
O\equiv U^{*}(p)-U^{*}(p_{c})=(p_{c}-p  )^{\beta }; \ \beta =\frac{1}{2}. 
\end{equation}

To study the dynamics away from criticality ($p \to p_{c}$
from below), we replace the recursion relation (1) by a differential
equation 
\begin{equation*}
-\frac{dU}{dt}=\frac{U^{2}-U+p }{U}.
\end{equation*}
Close to the fixed point we write \( U_{t}(p )=U^{*}(p ) \)
+ \( \epsilon  \) (where \( \epsilon \rightarrow 0 \)). This
gives  
\begin{equation}
\label{cc:epsilon}
\epsilon =U_{t}(p )-U^{*}(p )\approx \exp (-t/\tau ),
\end{equation}
where \( \tau =\frac{1}{2}\left[ \frac{1}{2}(p_{c}-p )^{-1/2}+1\right]  \).
Thus, near the critical point (for jamming transition) we can write
\begin{equation}
\tau \propto (p_{c}-p )^{-\alpha }; \ \alpha =\frac{1}{2}.
\end{equation}
Therefore the relaxation time diverges following a power-law as $p \to p_{c}$
from below.

One can also consider the breakdown susceptibility \( \chi  \), defined
as the change of \( U^{*}(p ) \) due to an infinitesimal increment
of the traffic stress \( p \)
\begin{equation}
\chi =\left| \frac{dU^{*}(p )}{d p }\right| 
=\frac{1}{2}(p_{c}- p )^{-\gamma };\gamma =\frac{1}{2}. 
\end{equation}
Hence the susceptibility diverges as
the average crowd level \( p\) approaches the critical value 
\( p_{c}=\frac{1}{4} \).

\subsubsection {At \(p  =p_{c} \)}
At the critical point (\( p =p_{c} \)), we observe
a different dynamic critical behavior in the relaxation of the failure process.
From the recursion relation (\ref{cc:recur}), it can be shown
that decay of the fraction \( U_{t}(p_{c}) \) of restaurants 
that remain in service at time \( t \) follows a simple power-law decay:
\begin{equation}
U_{t}=\frac{1}{2}(1+\frac{1}{t+1}),
\end{equation}
starting from \( U_{0}=1 \). For large \( t \) (\( t \to \infty  \)),
this reduces to \( U_{t}-1/2\propto t^{-\delta } \); \( \delta =1 \);
a power law, and is a robust characterization of the critical
state.

\subsection{Universality Class of the Model}

The universality class of the model can be checked [4] 
taking
two other types of restaurant capacity distributions $\rho(p)$: (I) linearly 
increasing
density distribution and (II) linearly decreasing density distribution
of the crowd fraction thresholds $p$ within the  limit $0$ and $1$. One can 
show that while 
$p_{c}$ changes with different strength distributions 
($p_c= \sqrt{4/27}$ for case (I) and $p_c=4/27$ for case (II), 
the critical behavior remains unchanged: $\alpha =1/2=\beta =\gamma$, 
$\delta =1$ for all these equal crowd or load sharing models.

\subsection {Fluctuation or Avalanche Statistics}

For $p < p_c$, the avalance size $m$ can be defined as
the fraction of restaurants falling out of service for an infinitesimal 
increase in the global crowd level (c.f. \cite{BKC:Pradhan:2002})
\begin{equation}
\label{cc:dmdsigma}
m = \frac{d M}{d p}; \ M = 1 -U^*(p).
\end{equation}
With $U^*(p)$ taken from (2), we get 
\begin{equation*}
 p_c - p \sim m^{-2}.
\end{equation*}
If we now denote
the avalanche size distribution by $P(m)$,
then $P(m)\Delta m$ measures $\Delta p$ along the $m$ versus $p$ curve in
(\ref{cc:dmdsigma}). In other words~\cite{BKC:Pradhan:2002,BKC:Pradhan:2007} 
\begin{equation}
\label{cc:dpdm}
P(m) = \frac{dp}{dm} \sim m^{-\eta}, \ \eta = 3.
\end{equation}

\section{Avalanche Dynamics: Finite $T$}

The above results are for $T \to \infty$, i.e, when any restaurant
fails to satisfy its customers, it falls out of business, and customers
never come back to it.
This would also be valid if $T$ is greater than the relaxation time
$\tau$ defined in (\ref{cc:epsilon}).
However, if such a restaurant again comes back to service 
(people either forget
or hope that it has got better in service and start choosing it
again) after $T$ evenings, then several scenerios can emerge.

If $T$ is very small, the recursion relation (\ref{cc:recur})
can be written as
\begin{equation}
\label{cc:recurTsm}
U_{t+1}=1-\frac{p}{1 - U_{t-T} + U_t}.
\end{equation}
In fact, if $T$ is large, $U_{t-T} = 1$ and (\ref{cc:recurTsm})
reduces to (\ref{cc:recur}). However, if $T$ is very small, say
of the order of unity, then at the fixed point,
$U_{t+1} = U^* = U^*_{t-T} = U_t$ and one gets
\[ U^* = 1-p.\]
From (\ref{cc:dmdsigma}) one gets
\begin{equation}
\label{cc:Pm}
P(m) \sim \delta(m-m_0)
\end{equation}
where $m_0$ is determined by the initial conditions.

\section{Conclusions}
We generalise here the El Farol bar problem to that of many bars. 
This leads us to the Kolkata restaurant problem, where the decision to go
to any restaurant or not is much simpler; it still depends on the previous
experience of course, as in the El Farol bar problem, but does
not explicitly depend on the memory size or the size of the
strategy pool. Rather, people getting disappointed with any restaurant
on any evening avoids that restaurant for the next $T$ evenings. This
generalised problem can be exactly analysed in some limiting
cases discussed here. In the $T \to \infty$ limit,
the fluctuation in the restaurant service can be shown to have
precisely an inverse cubic behavior (see eqn.~(\ref{cc:dpdm})), 
as widely seen in the stock market fluctuations. For very small 
values of $T$, the fluctuatuation distribution become $\delta$-function like
(see eqn.~(\ref{cc:Pm})).
For large but finite $T$, the system of restaurants in Kolkata will
survive at a (self-organised) critical state~\cite{cc:ChatChakunpub}.

\subsection*{Acknowledgement}

I am grateful to Arnab Chatterjee for his criticisms, discussions
and help in finalising the manuscript.

\end{document}